\documentclass[twoside,reqno]{HERON}
\usepackage{epsfig,cite,colordvi}
\usepackage{graphicx}
\usepackage{url}
\usepackage{amssymb,amsmath,amscd,epsf}
\usepackage{times}
\usepackage{makeidx}
\usepackage[english]{babel}
\begin{document}

\title{What could be learned about phase transitions, meson and nucleon structure in hot medium from a chiral quark-meson theory?}

\runningheads{Christo V. Christov}{What could be learned about phase transitions, meson and nucleon structure in medium... }

\begin{start}

\author{Christo V. Christov}{1,2}

\index{Christov, Chr. V.}

\address{1Institüt für Threoretische Physik II, Ruhr-Universität Bochum, Germany}{1}

\address{2Institute for Nuclear Research and Nuclear Energy, Bulgarian Academy of Sciences, Sofia 1784, Bulgaria}{2}

\begin{Abstract}
In this contribution I summarize and discuss the results of the bulk thermodynamic characteristics,  meson and nucleon structure in hot matter obtained in the framework of a chiral quark-meson theory. A hybrid NJL model  is used in which a Dirac sea of quarks is combined with a Fermi sea of quarks or of nucleons. In the model mesons are described as collective $\bar qq$ excitations and the nucleon appears as a baryon-number-one soliton of $N_c$ valence quarks coupled to both Dirac and Fermi sea. According to the model at some critical density and/or temperature phase transitions from nucleons to quarks as well as from Goldstone to Wigner phase are expected. At finite density the chiral order parameter and the constituent quark mass have a non-monotonic temperature dependence - at temperatures not close to the critical one they are less affected than in cold matter. The quark matter is rather soft against thermal fluctuations and the corresponding chiral phase transition is smooth. The nucleon matter is much stiffer and the phase transition is very sharp. In the case of quark matter a first-order transition is suggested at low temperatures ($T < 80$ MeV) which changes to a second-order one at higher temperatures. In contrast to the quark matter in the case of nucleon matter the thermodynamic variables show large discontinuities which is a clear indication for a first-order phase transition. In hot medium at intermediate temperature the nucleon soliton is more bound and less swelled than in the case of cold matter. At some critical temperature, which for nucleon matter coincides with the critical temperature for the phase transition, no more a localized solution is found. According to this model scenario one should expect a first-order phase transition from nucleon to quark matter. The results show that the hybrid model provides a consistent picture where the Fermi sea of quarks is reasonable only for hot matter with temperature and/or density around/above the critical values.
\end{Abstract}
\end{start}

\section{Introduction}
\begin{figure}[h]
\centering
\includegraphics[scale=0.8]{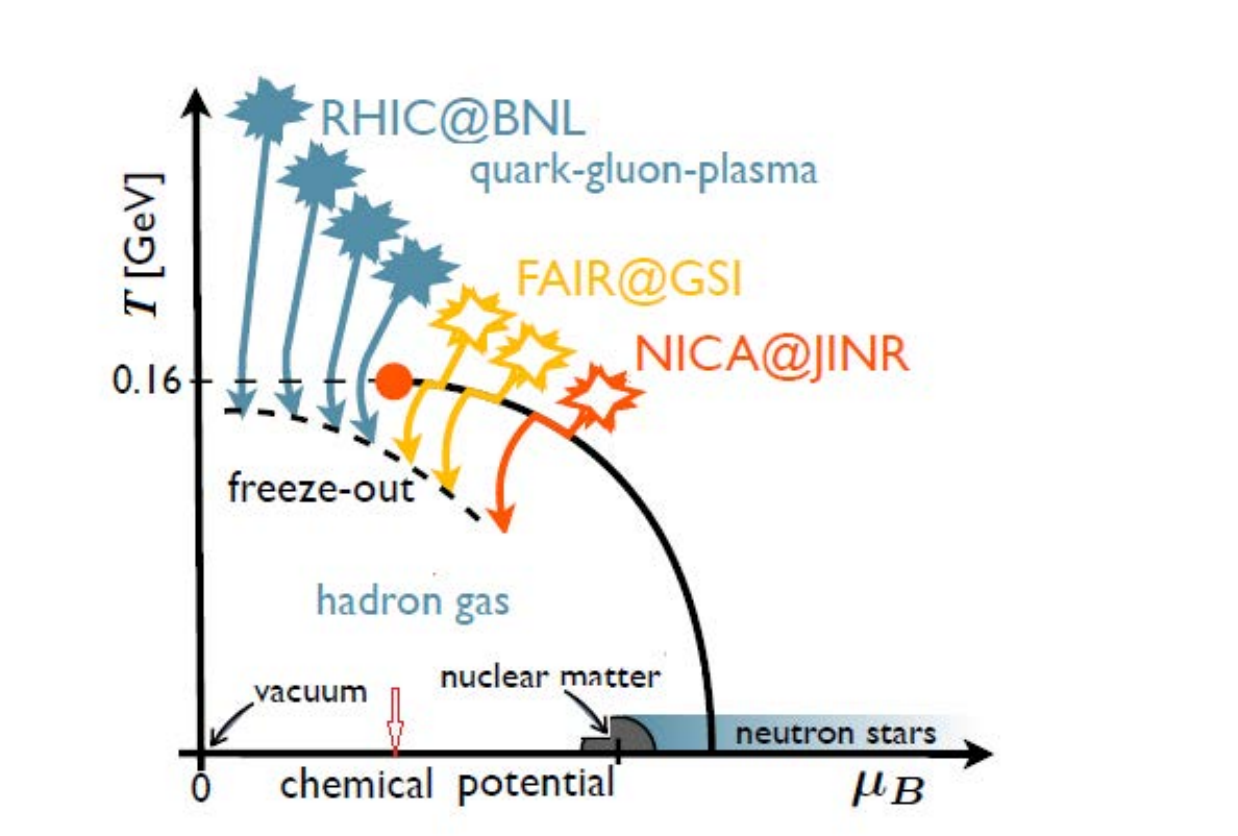}
\caption[]{Generally expected phase-diagram of QCD for nuclear matter. The solid lines show the
phase boundaries for different phases. The red solid circle corresponds to the estimated critical
point. The areas reached at different accelerator facilities are also shown. \cite{myref1}}\label{f01}
\end{figure}

At some finite density and/or temperature one generally expects a restoration of the chiral symmetry and a deconfinement, and hence a change of the structure of the hadrons immersed in a hot and dense medium. The bulk properties of hot nuclear and quark matter and especially the phase diagram of QCD is a topic of increasing interest since it is related to the evolution of the early universe after the Big Bang as well as to the processes in the interior of neutron stars. Rather encouragingly, direct experimental studies of such phenomena are now possible in the ultra-relativistic heavy-ion reactions accomplished at high-energy accelerators like RHIC, NA61 at the SPS as well as FAIR and NICA (see e.g. \cite{myref1} and for a review \cite{myref2}). It is illustrated on Figure~\ref{f01} where the phase diagram of QCD is schematically shown.

Obviously a detailed understanding of this phase diagram is one of the contemporary challenges of both particle and nuclear physics. In particular, the order of the phase transition is still an open question. It seems to be generally accepted that below some critical point there is a first order-phase transition which above the critical point undergoes a crossover transition to a second-order one. Around the critical point one expect a mix of the two phases with different order parameters (see e.g. for an overview \cite{myref3} and also \cite{myref4}). 

From the side of theory these expected low critical values mean that we have to deal with a non-perturbative phenomena. So far the analytical as well as the numerical (lattice QCD) methods (see review \cite{myref2} and references therein) are still not developed enough to allow the solution of low-energy non-perturbative phenomena, especially if nucleons are involved. It is the reason for applying in the last two decades different effective models like e.g. NJL \cite{myref8} and linear sigma model \cite{myref6} to study these phenomena. These models are rather attractive since although lacking confinement they incorporate, similar to QCD, the chiral symmetry and also allows for its spontaneous breaking. It was the motivation to apply a chiral quark-meson model (see for review \cite{myref4} and references therein) for these analyses.  The model is able to reproduce quite reasonably the nucleon and delta properties as well as the corresponding formfactors in vacuum. Within the model the nucleon in vacuum  is considered as a bound state of N, quarks coupled to the polarized Dirac sea. The mean-field approximation is used, which means that the meson quantum (loop) effects are not considered. Since the meson loop effects are dominant at low temperatures and vanishing density (pions are the lightest mode) the results are restricted mainly to the case of finite density and relatively large temperatures where the nucleon and quark degrees of freedom are most relevant. In order to consider the medium effects in the model a Fermi sea is added.

\section{Quark-meson chiral model at finite temperature and density}

The SU(2)-version of the NJL Lagrangian \cite{myref8} contains chirally invariant local scalar and pseudoscalar four-quark
interaction:

\begin{equation}
\label{myeq1}
\mathcal {L} =\bar\Psi (i\not\partial - \it m_0)\Psi + \dfrac{1}{ 2}  \it G [(\bar\Psi\Psi)^2 + (\bar\Psi\it i\mathbf\tau\gamma_5\Psi)^2]
\, ,
\end{equation}

where $\Psi$ is the quark field, $\it G$ is the coupling constant, $\mathbf\tau$ are the Pauli matrices in the
isospin space and $m_0$ is the current quark mass taken equal for both up and down quarks.
Applying the well-known bosonization procedure \cite{myref9} the NJL model is expressed in
terms of the auxiliary meson fields $\sigma, \pi$:

\begin{equation}
\label{myeq2}
\mathcal {L} =\bar\Psi (i\not\partial - \sigma -i\mathbf\pi\cdot\mathbf{\tau}\gamma_5)\Psi - \dfrac{1} {2G}(\sigma^2+\mathbf\pi^2) + \dfrac{m_0} {G}\sigma
\, ,
\end{equation}

Using the functional integral technique the quantized theory at finite temperature and
density can be written in terms of the corresponding euclidean grand canonical partition
function \cite{myref10}:

\begin{equation}
\label{myeq3}
\begin{split}
\mathcal {Z} &= \mathbf {Tr} \exp \{-\beta (\it H - \mu \it N\} \\ &= \dfrac{1} {Z_0} \int \mathcal D\Psi\ \mathcal D \Psi^\dagger  \mathsf {exp} \{\int_0^\beta \mathsf d\tau \int_V\mathsf d^3 x (\mathcal {L}-\Psi^\dagger \mu\Psi)\}
\, ,
\end{split}
\end{equation}

where $\it V$ is the volume of the system, $\beta$ is the inverse temperature and $\mu$ is the chemical
potential. The integration over the quarks can be done exactly, whereas for the integration over the mesons we use a large $N_c$ saddle-point (mean-field) approximation. This means that the meson fields are treated classically - no meson loops are taken into account. Following  \cite{myref10} the integration over the imaginary time is replaced by a sum over fermionic Matsubara frequencies. Finally the effective action is expressed as

\begin{equation}
\label{myeq4}
\begin{split}
S_{eff}(\mu,\beta) = - \ln \mathcal {Z} = - \beta V N_c \sum_\alpha \lbrace \dfrac{1} {2}(\epsilon_\alpha - \mu) + \dfrac{1} {\beta} \ln[1 +\mathsf e^{-(\epsilon_\alpha - \mu)\beta}]\rbrace \\ + \beta\int_V \mathsf {d^3} x [\dfrac{1} {2G} (\sigma^2+\mathbf\pi^2) - \dfrac{m_0} {G}\sigma]
\, .
\end{split}
\end{equation}

The energies $\epsilon_\alpha$ are eigenvalues of the one-particle hamiltonian $h$.

\begin{equation}
\label{myeq5}
h\Phi_n \equiv [\dfrac{\mathbf{\alpha} {\cdot} \mathbf{\Delta}}{i} + \gamma_0(\sigma + i\mathbf\pi\cdot\mathbf\tau\gamma_5)] \Phi_n = \epsilon_n\Phi_n \, ,
\end{equation}

and $\Phi_n$ are eigenfunctions. The saddle-point solution makes the effective action stationary

\begin{equation}
\label{myeq6}
\dfrac{\partial S_{eff}}{\partial \sigma} \lvert_{\sigma_c} = \dfrac{\partial S_{eff}}{\partial \mathbf\pi} \lvert_{\mathbf\pi_c} = 0 \, ,
\end{equation}

The number of particles $N$ in the volume $V$ is kept fixed

\begin{equation}
\label{myeq7}
N = - \dfrac{1}{\beta}\dfrac{\partial S_{eff}}{\partial \mu} \lvert_{\sigma_c , \mathbf\pi_c}  \, .
\end{equation}

Here $\sigma_c$ and $\mathbf\pi_c$ are the “classical” values of the meson fields and $\mu$ is the
chemical potential related to the number of particles $N$.

The thermodynamic characteristics of a many-body system are specified by the
thermodynamic potential

\begin{equation}
\label{myeq8}
\Omega(\mu,\beta) \equiv \dfrac{S_{eff}(\mu,\beta)}{\beta V}  \, .
\end{equation}

It should be noticed that the saddle-point solution ($\sigma_c$ , $\mathbf\pi_c$) minimizes not $\Omega$ but the
Helmholtz free energy

\begin{equation}
\label{myeq9}
F = \Omega - \mu \dfrac{\partial\Omega}{\partial\mu}  \, 
\end{equation}

with a constraint (\ref{myeq7}) and $\mu$ playing a role of a Lagrange multiplier.

In the mean-field approximation (leading order in $\tfrac {1}{N_c}$ the inverse meson propagator
is given by the second variation of the effective action at the stationary point ($\sigma_c$ , $\mathbf{\pi_c}$):

\begin{equation}
\label{myeq10}
K_\phi^{-1}(x-y) =  \dfrac{\partial^2S_{eff}}{\partial\phi(x)\partial\phi(y)}\lvert_{\phi_c}  \, .
\end{equation}

The on-shell meson masses correspond to the poles of the meson propagator at $q = 0$. The physical quark meson coupling constants are given by the residue of the propagator at the pole

\begin{equation}
\label{myeq11}
g_\phi^{2} =  \lim_{q^2 \rightarrow {- m^2_\phi}} (q^2 + m^2_\phi) K_\phi(q^2) \, .
\end{equation}

Due to the local four-fermion interaction the lagrangian (\ref{myeq1}) is not renormalizable and a regularization procedure with an appropriate cut-off $\Lambda$ is needed to make the effective action finite:

\begin{equation}
\label{myeq12}
\mathbf {Tr} \ln \hat{A} \rightarrow - \mathbf {Tr} \int_{\Lambda^{-2}}^\infty \dfrac{\mathsf{d} s} {s}  \mathbf {e}^{-s\hat{A}}\, .
\end{equation}

Actually only the part of the effective action $S_{eff}(\mu=0,\beta=\infty)$ coming from the Dirac sea (negative-energy part of the spectrum), is divergent and one needs to regularize it. Here the proper-time regularization scheme is used. The difference $S_{eff}(\mu,\beta) - S_{eff}(\mu=0,\beta=\infty)$ is finite (Fermi sea contribution) and does not need any regularization. Moreover, any regularization of the medium part would suppress the medium effects (\cite{myref11}), since the positive part of the spectrum would be affected by the cutoff as well. 

Two different scenario are considered: Fermi sea of quarks as well as Fermi sea of nucleons. In both cases the Dirac sea consists of quarks and it determines the vacuum sector. In both pictures the mesons appear as $\hat{q} q$ excitations but they are also directly coupled to the Fermi sea as well. In the case of Fermi sea of quarks at finite temperature/density single quarks are allowed to be excited leaving holes (antiquarks) in the Dirac sea. Apparently because of confinement this picture is applicable only at temperature/density above the critical ones - after the phase transition.  

Since the Fermi sea contribution is finite it is straightforward to write the Fermi sea part in terms of nucleon.

\begin{equation}
\label{myeq13}
S_{med}^N(\mu,\beta) = \sum_{\epsilon_\alpha^N<0} \lbrace (\mu^N - \epsilon_\alpha^N) - \dfrac{1} {\beta} \ln[1 +\mathsf e^{-\beta (\epsilon_\alpha^N - \mu^N)}]\rbrace
\, .
\end{equation}

The energies $\epsilon_\alpha^N$ are the solutions of the corresponding Dirac equation.

\begin{equation}
\label{myeq14}
h^N\Phi_n^N \equiv [\dfrac{\mathbf{\alpha} {\cdot} \mathbf{\Delta}}{i} + \beta g_N(\sigma + i\mathbf\pi\cdot\mathbf\tau\gamma_5)] \Phi_n^N = \epsilon_n^N\Phi_n^N \, ,
\end{equation}

The meson fields are coupled to the nucleons with a coupling constant $g_N$ which relates the nucleon mass to the non-zero expectation value of the scalar meson field (constituent quarks mass $M_0$) in vacuum.

\begin{equation}
\label{myeq15}
M_N = g_N M_0\, .
\end{equation}

As in the quark spectrum, there is a gap of $2M_N$ in the nucleon spectrum which separates the negative part of the spectrum from the positive one.
The chemical potential $\mu_N$ is fixed by the baryon number (\ref{myeq7}).

\section{Fixing the model parameters}

In vacuum ($\mu = 0, T = 0$) the stationary conditions lead to a transitionally invariant solution $\sigma_c = M_0$ and $\mathbf\pi_c = 0$.

The parameters of the model, namely the current mass $m_o$, the cutoff $\Lambda$ and the coupling constant $G$ In the vacuum sector, are fixed reproducing the physical pion mass $m_\pi = 140$ MeV and the pion decay constant $f_\pi = 93$ MeV.
It leads to the well-known Goldberger-Treiman (GT) relation on the quark level

\begin{equation}
\label{myeq16}
M_0 = g_\pi f_\pi\, .
\end{equation}

and one also recovers the Gell-Mann-Oakes-Renner (GMOR) relation:

\begin{equation}
\label{myeq17}
m_\pi^2 = -m_0\langle\bar\Psi\Psi\rangle + O(m_0^2) \,.
\end{equation}

The only free parameter remained is the constituent quark mass $M_0 = 420$ MeV is taken to reproduce properly the properties of a free nucleon (see \cite{myref4} and references therein). The corresponding value of the proper-time cutoff used is $\Lambda = 640$ MeV. The value is taken large enough not to influence to results.

\section{Phase transition and meson properties at finite temperature and density}

\begin{figure}[h]
\centering
\includegraphics[scale=0.8]{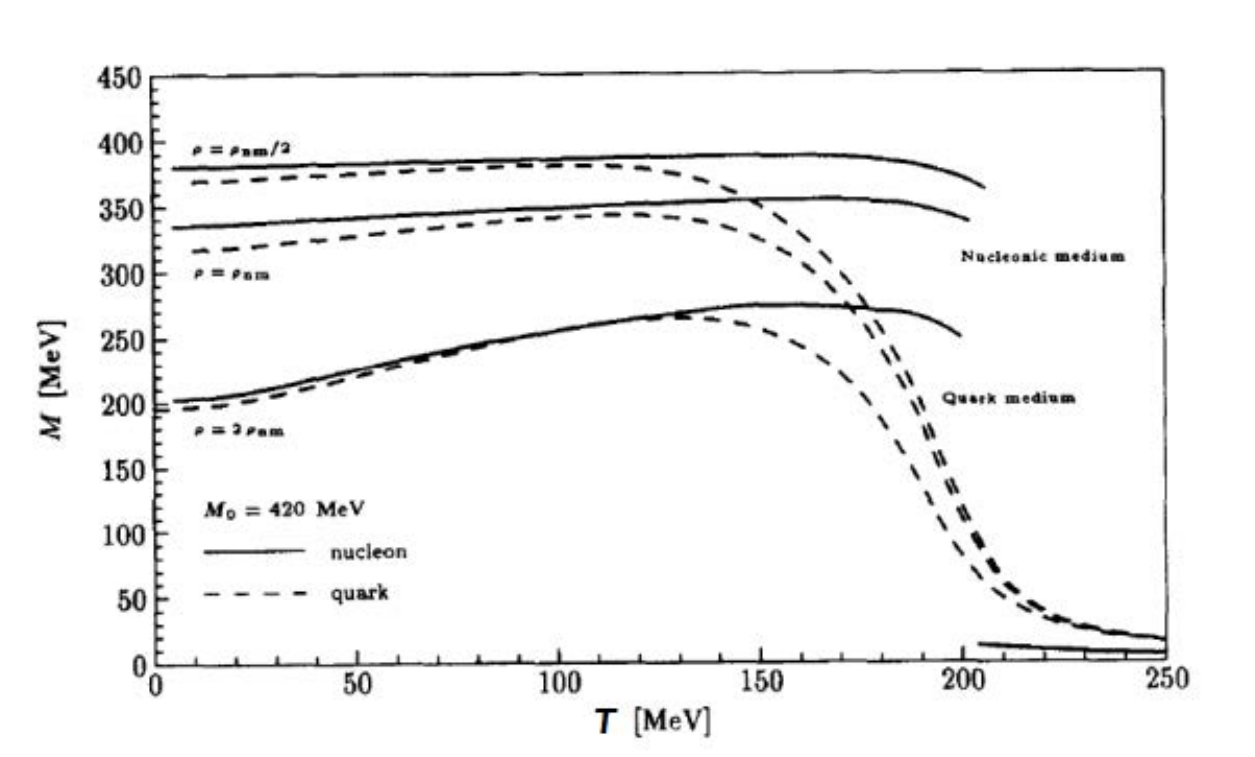}
\caption[]{Constituent quark mass M as a function of temperature for different densities in a quark (dashed lines)
and a nucleon medium (solid lines). At both vanishing temperature and density it is fixed to $M_0 = 420$ MeV}\label{f02}
\end{figure}

\begin{figure}[h]
\centering
\includegraphics[scale=0.8]{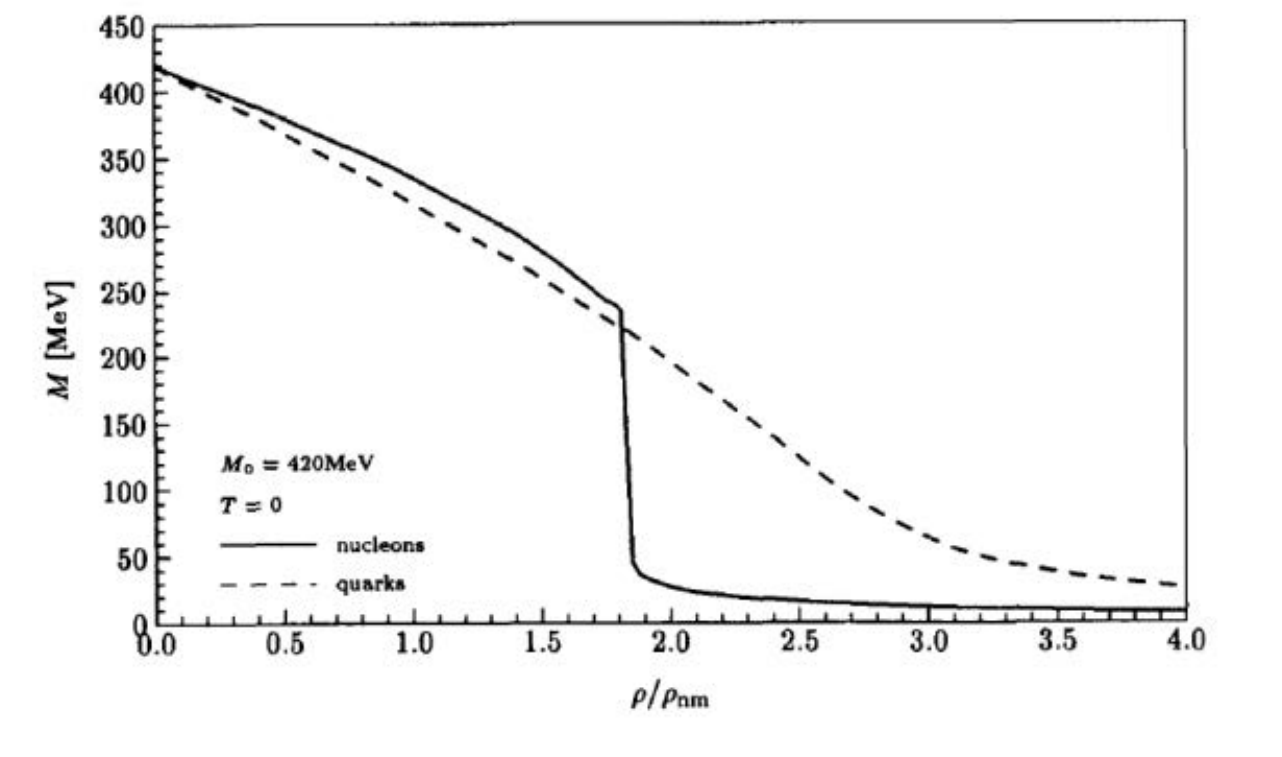}
\caption[]{Constituent quark mass M as a function of density for different temperatures.}\label{f03}
\end{figure}

The results \cite{myref12} for the constituent mass M as a function of temperature and density and are presented in Figure~\ref{f02} and Figure~\ref{f03}. At both vanishing temperature and density it is fixed to $M_0 = 420$ MeV

At finite density the constituent mass $M$ is a non-monotonic function of temperature. It means that in hot matter the mass is less affected by the medium compared to the case of cold matter. At low $T$ values the curves are close to those of quark matter.  However already at intermediate temperatures $T > 120$ MeV they start to deviate significantly. At some critical values of temperature and/or density the constituent mass $M$ is reduced to the current mass $m_0$ which is an indication for a transition from Goldstone to the Wigner phase where the chiral symmetry is not spontaneously broken. Whereas the quark matter is quite soft against thermal fluctuations, the nucleon matter is much stiffer and the corresponding chiral phase transition is rather sharp. In order to determine the order of the phase transition we follow the Ehrenfest classification. Both the chiral condensate $\langle\bar\Psi\Psi\rangle$ and the constituent quark mass show a discontinuity at the critical temperature of about 200 MeV. In fact, the system jumps between two minima. The latter suggests that in the case of nucleon matter we have to deal with a first order phase transition even in case of vanishing density and high temperature in contrast to the case of quark matter. The results of  $M$ at $T = 0$ and finite $\rho$ suggest a critical density ($\rho_c \approx 2 \rho_{nm}$ with $\rho_{nm} = 0.16 fm^3$.) which in fact, as we will see later, is also the critical density for the delocalization of the soliton in cold nucleon matter.The same is valid for the  temperature $T_c$  which makes this model picture consistent. 

\begin{figure}[h]
\centering
\includegraphics[scale=0.8]{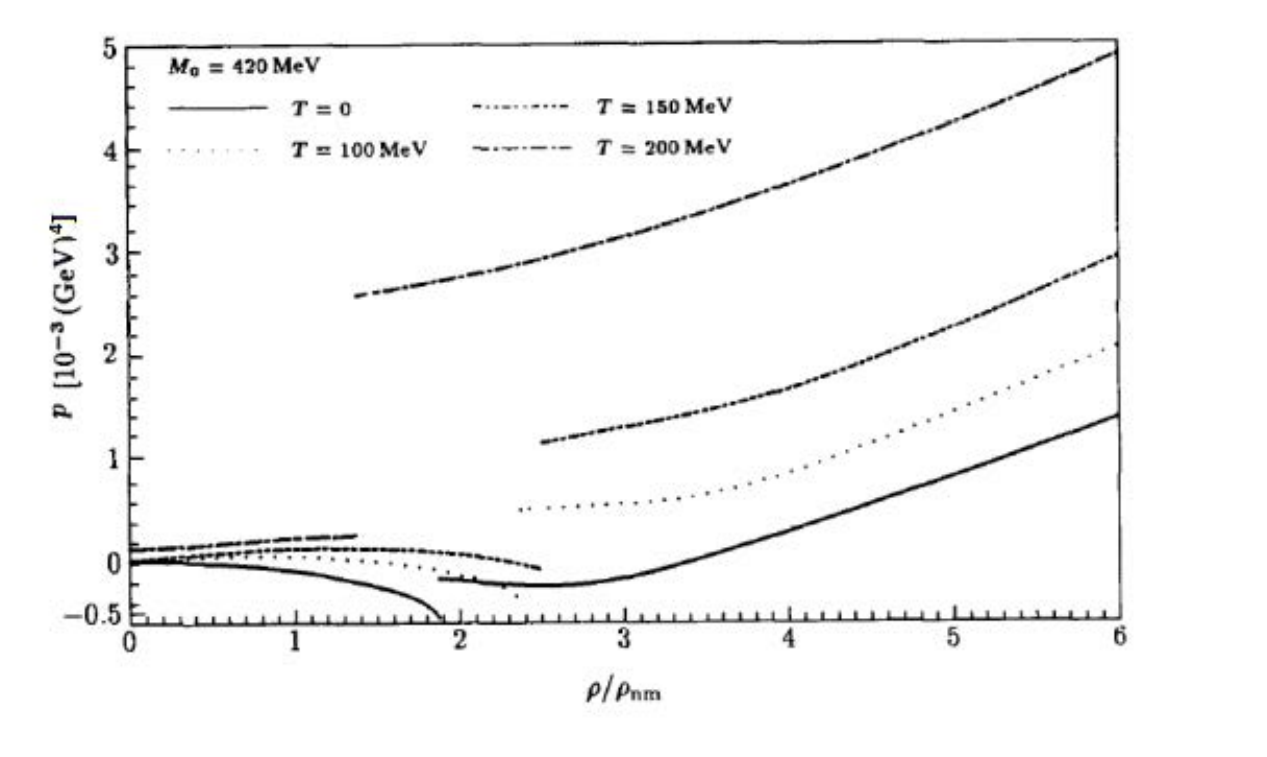}
\caption[]{EOS (pressure versus density) for different temperatures. The pressure of the vacuum is subtracted.
The discontinuities correspond to a phase transition from nucleon to quark matter.}\label{f04}
\end{figure}

\begin{figure}[h]
\centering
\includegraphics[scale=0.8]{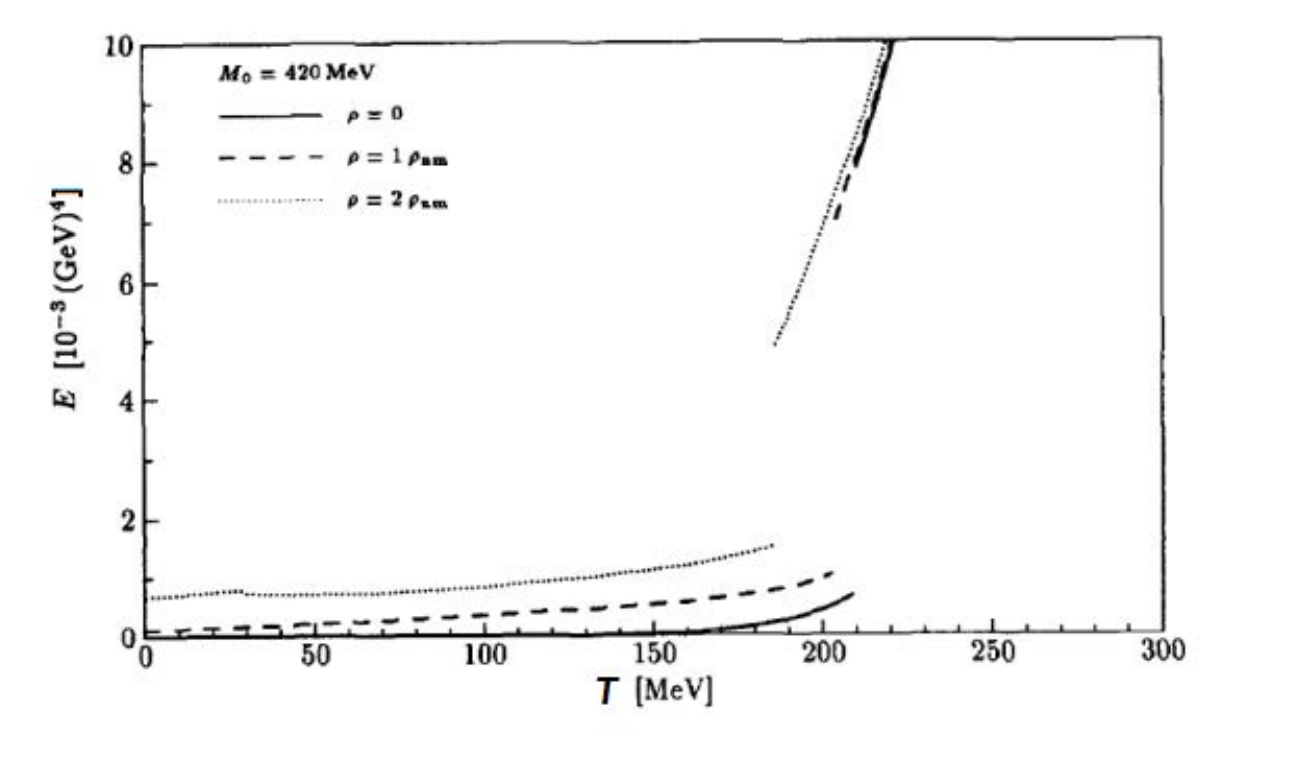}
\caption[]{The same as Figure~\ref{f04} but for the energy density. The discontinuities correspond to a phase transition from nucleon to quark matter.}\label{f06}
\end{figure}

According to the present model picture, at some critical temperature and/or density one expects a chiral phase transition together with a delocalization transition from nucleon matter to quark matter. It can be clearly seen from the corresponding EOS (pressure versus baryon density) for different temperatures plotted in Figure~\ref{f04}. In this Figure we combine the results from nucleon matter (below the critical temperatures) in the hybrid model with those of quark matter after the transition. Since we do not include vector mesons in the hybrid model, we are not able to reproduce the nuclear matter saturation at zero temperature and finite density which in the Walecka approach \cite{myref13} is due to the interplay between the $\sigma$-meson attraction and the $\omega$-meson repulsion. All curves in Figure~\ref{f04} show rapid change discontinuity at the delocalization transitions from nucleon to quark matter. Similar behavior can be seen in Figure~\ref{f06} \cite{myref12} for the energy density as a function of $T$ for different densities.

\begin{figure}[h]
\centering
\includegraphics[scale=0.8]{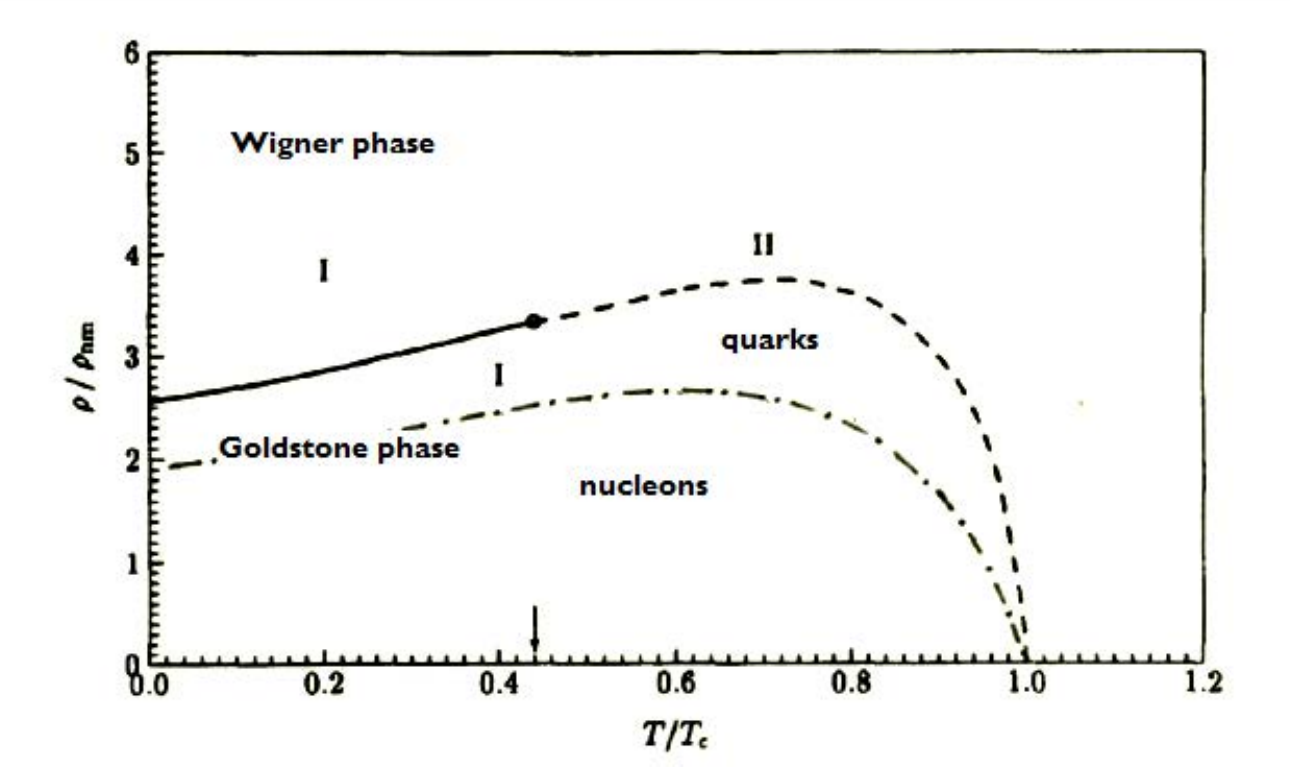}
\caption[]{$\rho-T$ critical chiral phase diagram for quark matter\cite{myref12}. The upper lines separates the Goldstone phase with the chiral symmetry breaking from the Wigner phase where the chiral symmetry is restored. The solid line shows the critical values for which there is a first-order transition whereas the dashed line corresponds to the second-order one. The  arrow shows the temperature at which the order is changed. The dash-dot line represents the $\rho-T$ critical phase diagram for transition from nucleon matter to quark one.}\label{f024}
\end{figure}

The phase diagrams for quark matter as well as for nucleon matter are shown in  Figure~\ref{f024} \cite{myref12}. As can be seen in quark matter at temperatures $T > 90$ MeV a second-order phase transition is predicted.  At lower temperatures in quark matter a first-order transition is expected. For the nuclen matter the model predicts always a first-order transition. It should be noted also that below the critical temperature $T_c$ the critical density $\rho_c$ for the phase  transition in nucleon matter is always smaller than one in quark matter. It means that at least according to the present model picture at temperatures below the critical $T_c$ two phase transitions are expected - a phase transition from nucleon to quark matter, where the chiral symmetry is still partially broken, followed by a phase transition from Goldstone to Wigner phase at a higher density in quark matter. 

Concerning the order of the phase transition the model picture suggests at low temperatures ( $< 90 $ MeV) a first-order transition from both the nucleon and from quark matter. At higher temperatures the picture is more complicated. For the phase transition from nucleon matter to quark one expects first-order whereas from quark matter the second-order phase transition is predicted. It means that at higher temperatures one could expect a mix of different phases with different order parameters. 

Apart from the medium part written in terms of nucleons the inverse meson propagators in the nucleon medium have the same structure (29) as in the quark medium. The meson masses in the medium defined as the lowest zero solution of the inverse meson propagators at q = 0 are presented in Figure~\ref{f04}.

\begin{figure}[h]
\centering
\includegraphics[scale=0.8]{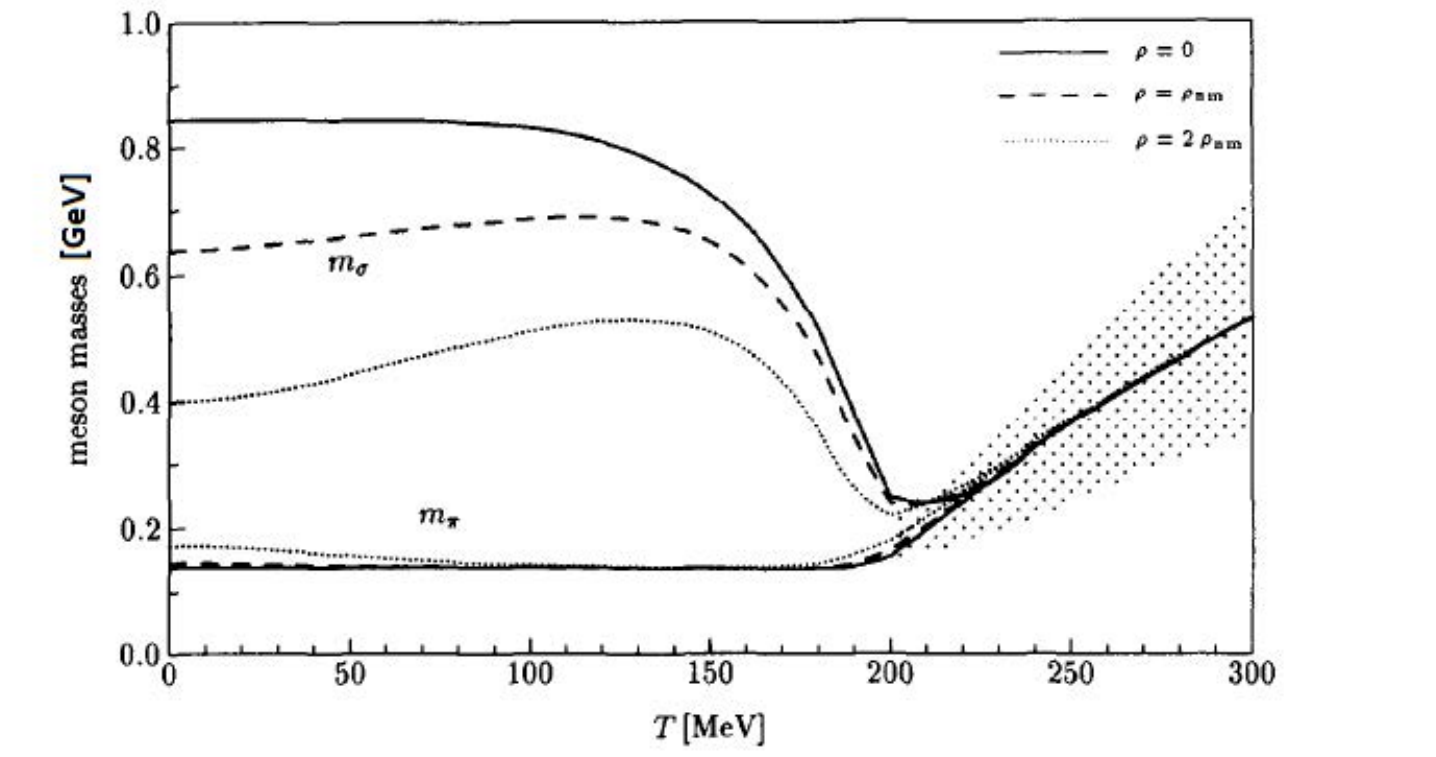}
\caption[]{Meson masses in a quark medium as a function of temperature for different densities. The shadowed area shows the width of the meson resonances.}\label{f05}
\end{figure}

In chiral limit the pions remain Goldstone bosons and their mass remains almost unchanged also in the medium, whereas the sigma mass $m_\sigma$ follows the constituent quark mass. Both GMOR and GT relations are also valid in the medium. Since in the Wigner phase the constituent quark mass $M$ vanishes up to the current mass $m_0$, the mesons become degenerated in mass and appear as parity-doubled mesons.In the Wigner phase, after the deconfinement transition (suggested also by the lattice results) the meson states appear as resonances in the continuum and the poles of the meson propagators are complex. In the nucleon medium similar to the constituent mass M the mass of the sigma meson has a discontinuity at the critical temperature.

\section{Nucleon as a non-topological $B = 1$ soliton in the medium}

The nucleon in a hot medium appears as B = 1 localized bound solution (soliton) of $N_c$ valence quarks interacting with Fermi and Dirac sea both getting polarized due to the interaction. The thermodynamic potential (effective action) includes an explicit valence quark contribution as well as Dirac and Fermi sea contributions:

\begin{equation}
\label{myeq18}
\begin{split}
S_{eff}(\mu,\beta) = N_c\Theta(\epsilon_{val})\epsilon_{val} + N_c \sum_{\epsilon_n < 0} \lbrace R^\Lambda_{\dfrac{3}{2}}(\epsilon_n)-R^\Lambda_{\dfrac{3}{2}}(\epsilon_n^0) + (\mu-\epsilon_n) \\ + \dfrac{N_c} {\beta} \ln[1 +\mathsf e^{-\beta(\epsilon_n - \mu)}]\rbrace + \dfrac{1}{V}\int_V \mathsf {d^3} x [\dfrac{1} {2G} (M^2-M_0^2) - \dfrac{m_0} {G}\sigma-M_0]
\, ,
\end{split}
\end{equation}

where $R^\Lambda_{\tfrac{3}{2}}(\epsilon)$ is the proper-time regularization function of the form (\ref{myeq12}).
The meson fields are assumed to be in a hedgehog form

\begin{equation}
\label{myeq19}
\sigma(\vec r) = \sigma(r) \; \text {and} \; \mathbf{\pi}(\vec r) = \hat r \pi (r) \, .
\end{equation}

restricted on the chiral circle.

\begin{equation}
\label{myeq20}
\sigma^2  + \mathbf{\pi}^2 = M^2 \, .
\end{equation}

We use a numerical self-consistent iterative procedure (see \cite{myref4} and references therein) solving in an iterative way the Dirac equation together with the equations of motion of the meson fields with the constraint $N_B$ = const. The latter is a condition fixing the chemical potential $\mu$. In the case of fixed $T$ and $\rho$ the proper way to describe the equilibrium state of a thermodynamic system is to use the free energy (\ref{myeq9}). Hence, the energy of the B = 1 soliton (effective soliton mass) is given by the change of the free energy when the $N_c$ valence quarks are added to the medium. Subtracting the free energy $F( \mu_0, ß)$ of the unperturbed Fermi and Dirac sea (translationally invariant medium solution), the soliton energy is given by the sum of the energy of the valence quarks and the contributions due to the polarization of both continua:

\begin{equation}
\label{myeq21}
\begin{split}
E_{sol} = N_c\eta_{val}\epsilon_{val} + N_c \sum_{\epsilon_n < 0} \lbrace R^\Lambda_{\dfrac{3}{2}}(\epsilon_n) + (\mu-\epsilon_n) + \dfrac{N_c} {\beta} \ln[1 +\mathsf e^{-\beta(\epsilon_n - \mu)}]\rbrace \\ + \dfrac{1}{V}\int_V \mathsf {d^3} x \dfrac{m_0} {G}\sigma +\mu N_c \rho_B -F(\mu_0,\beta)
\, ,
\end{split}
\end{equation}

where $\rho_B$ is the baryon density. Accordingly the soliton baryon density distribution is split in valence, sea and medium parts.

\begin{equation}
\label{myeq22}
\begin{split}
\rho_{sol} = N_c\theta(\epsilon_{val})\Phi^\dagger_{val}(\vec r)\Phi_{val}(\vec r) + \dfrac{1}{2}\sum_{\epsilon_n} \Phi^\dagger_{val}(\vec r)\Phi_{val}(\vec r)\mathbf{sgn}(-\epsilon_n)  \\ + \sum_{\epsilon_n > M} \dfrac{\Phi^\dagger_{val}(\vec r)\Phi_{val}(\vec r)} {1 +\mathsf e^{-\beta(\epsilon_n - \mu)}} - \sum_{\epsilon_n < 0} \dfrac{\Phi^\dagger_{val}(\vec r)\Phi_{val}(\vec r)} {1 +\mathsf e^{-\beta(\epsilon_n - \mu)}} - \rho_B
\, ,
\end{split}
\end{equation}

We do not find a localized solution (soliton) at temperatures larger than a critical
value $T_c \approx 200$ MeV. It means that at sufficiently large values the temperature effects
simply disorder the system and destroy the soliton. Within the present model picture this
may be interpreted as an indication for a delocalization of the nucleon in a hot medium.
However, one should keep in mind that the model lacks confinement. 

\begin{figure}[h]
\centering
\includegraphics[scale=0.8]{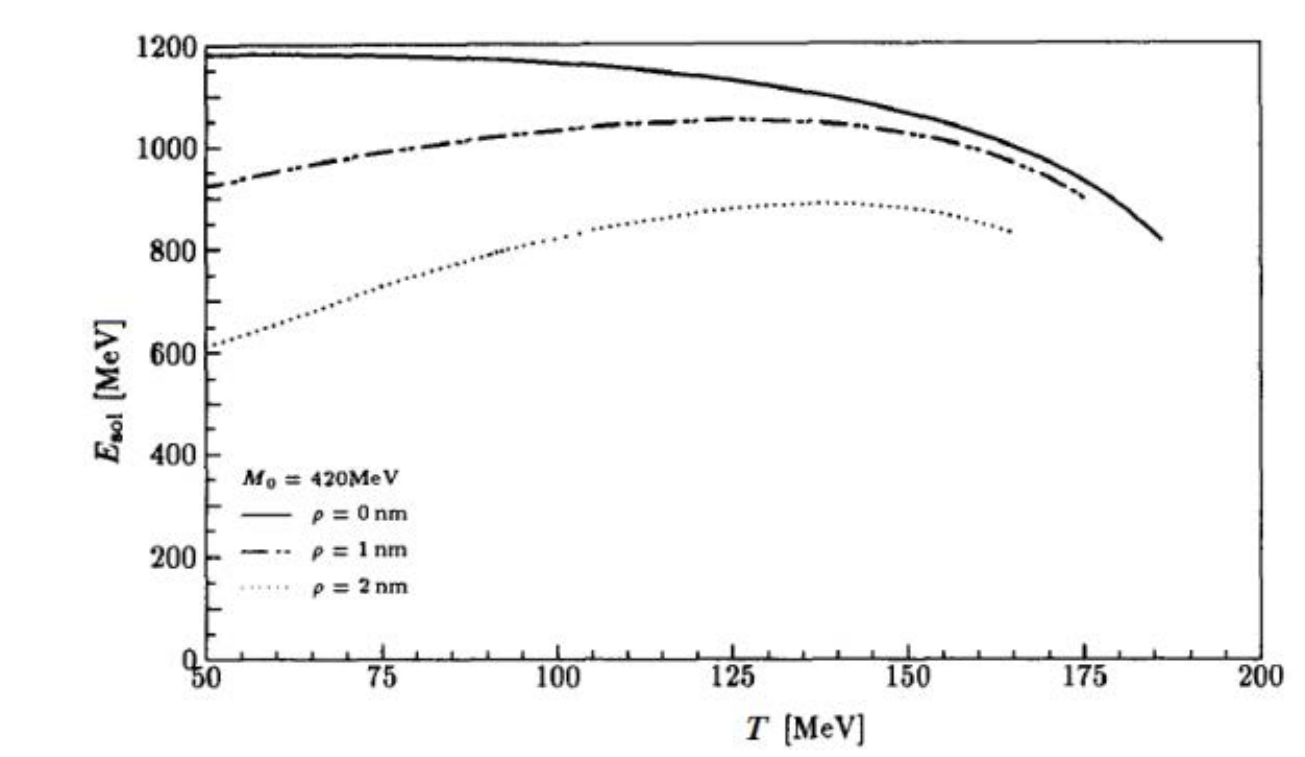}
\caption[]{B = 1 soliton energy as a function of the temperature at three different quark medium densities.}\label{f071}
\end{figure}

All detailed results can be found in \cite{myref12}. In Figure~\ref{f071} the temperature dependence of the calculated $B = 1$ soliton energy are presented for the different quark matter density. In case of finite $T$ and zero $\rho$ the soliton energy shows
almost no reduction for temperatures not close to the critical one, whereas in the
opposite case of finite $\rho$ and zero $T$ the soliton energy is linearly decreasing with the
density. In a hot medium (both $T$ and $\rho$ finite) the temperature clearly suppresses the
finite density effects and stabilizes the soliton. Further, at densities larger than two times
$\rho_{nm}$ the soliton exists only at intermediate temperatures 100 MeV $< T < 200$ MeV. It
means that according to the present model calculations, the soliton is more stable in hot
matter than in cold matter. In fact, this can be easily understood. Because of a gap in
the quark spectrum, $2M$, the Dirac sea is much less affected by the temperature than
the Fermi sea (positive-energy part of the spectrum). At temperatures close to the critical one,
the thermal fluctuations become comparable with the chiral order parameter (chiral
condensate $\langle\bar\Psi\Psi\rangle > 0$). They completely disorder the system and destroy the
soliton.

\begin{figure}[h]
\centering
\includegraphics[scale=0.8]{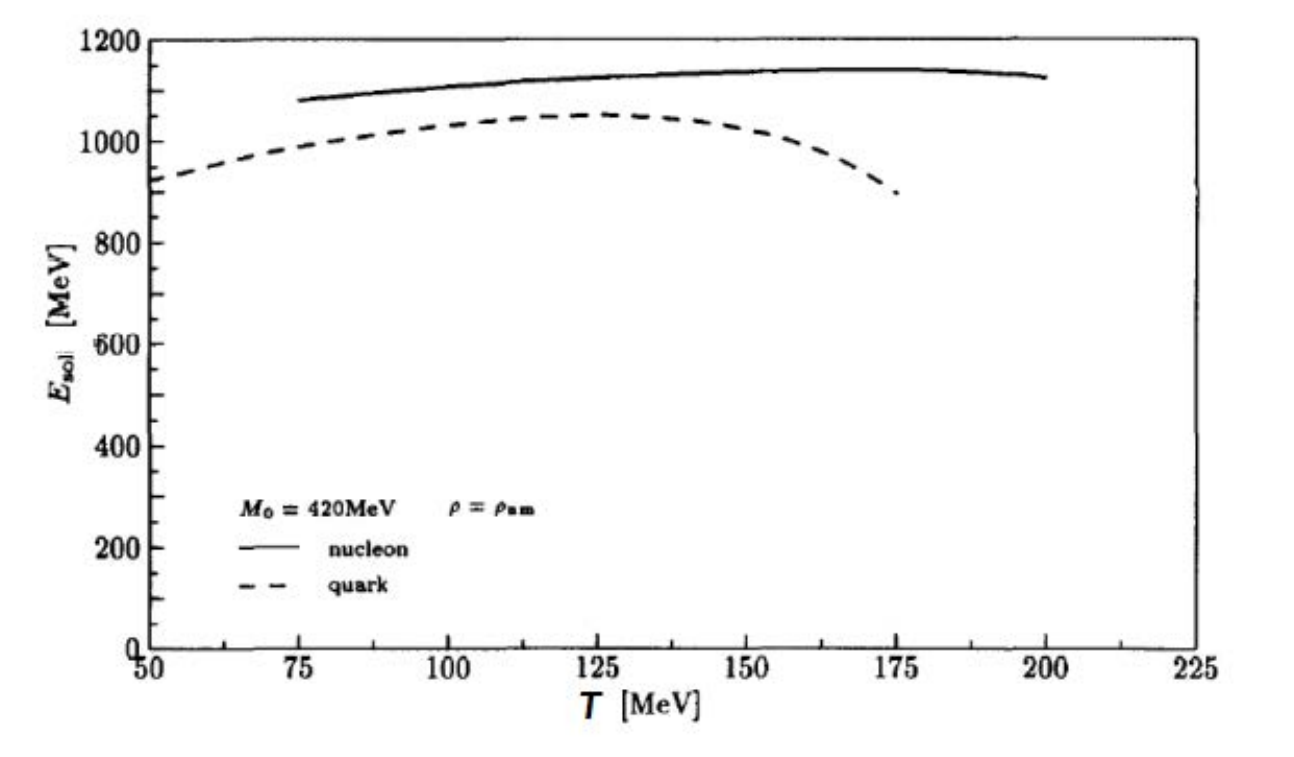}
\caption[]{B = 1 soliton energy in nucleon matter at $\rho = \rho_{nm}$ compared to those in quark matter.}\label{f07}
\end{figure}

In Figure~\ref{f07} the temperature dependence of the calculated $B = 1$ soliton energy for the nuclear matter density is compared with those of quark matter. The two curves have similar trends at intermediate values of $T$ and start to deviate at $T$ close to the
critical temperature $T_c$. The soliton in nucleon matter is more bound and less affected by the temperature. 

\begin{figure}[h]
\centering
\includegraphics[scale=0.8]{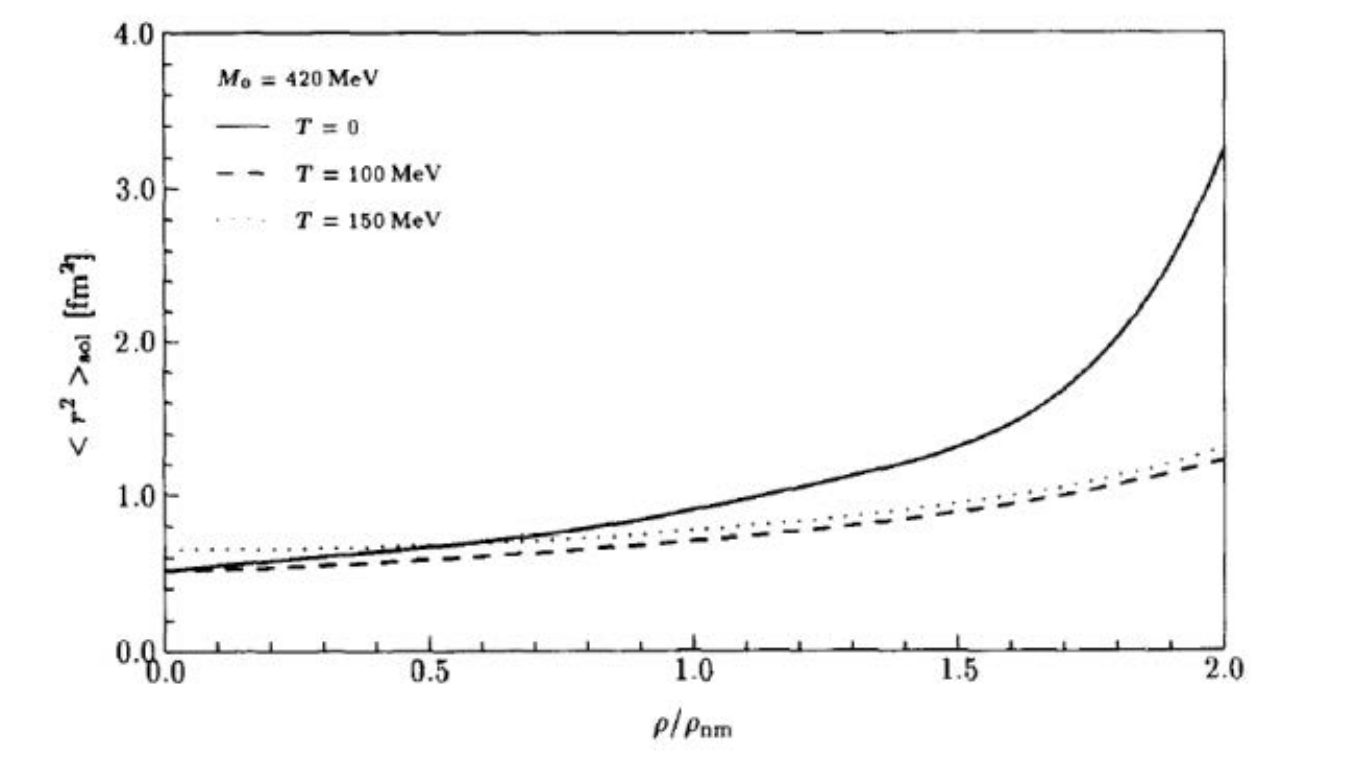}
\caption[]{B = 1 soliton m.s. radius as a function of density for vanishing \cite{myref14} as well as for finite temperature
values.}\label{f08}
\end{figure}

In order to illustrate the change of the soliton structure in a hot medium we also
plot the soliton square radius as a function of density for different temperatures
in Figure~\ref{f08}. All curves  show a clear trend to grow rapidly at temperatures
close to the critical values which is an indication for a delocalization of the soliton. At
both finite density and temperature, however, the radius is smaller than in the case of cold matter
which is a sign for stabilization of the soliton in a hot medium compared to the case
of a cold medium.

\begin{figure}[h]
\centering
\includegraphics[scale=0.8]{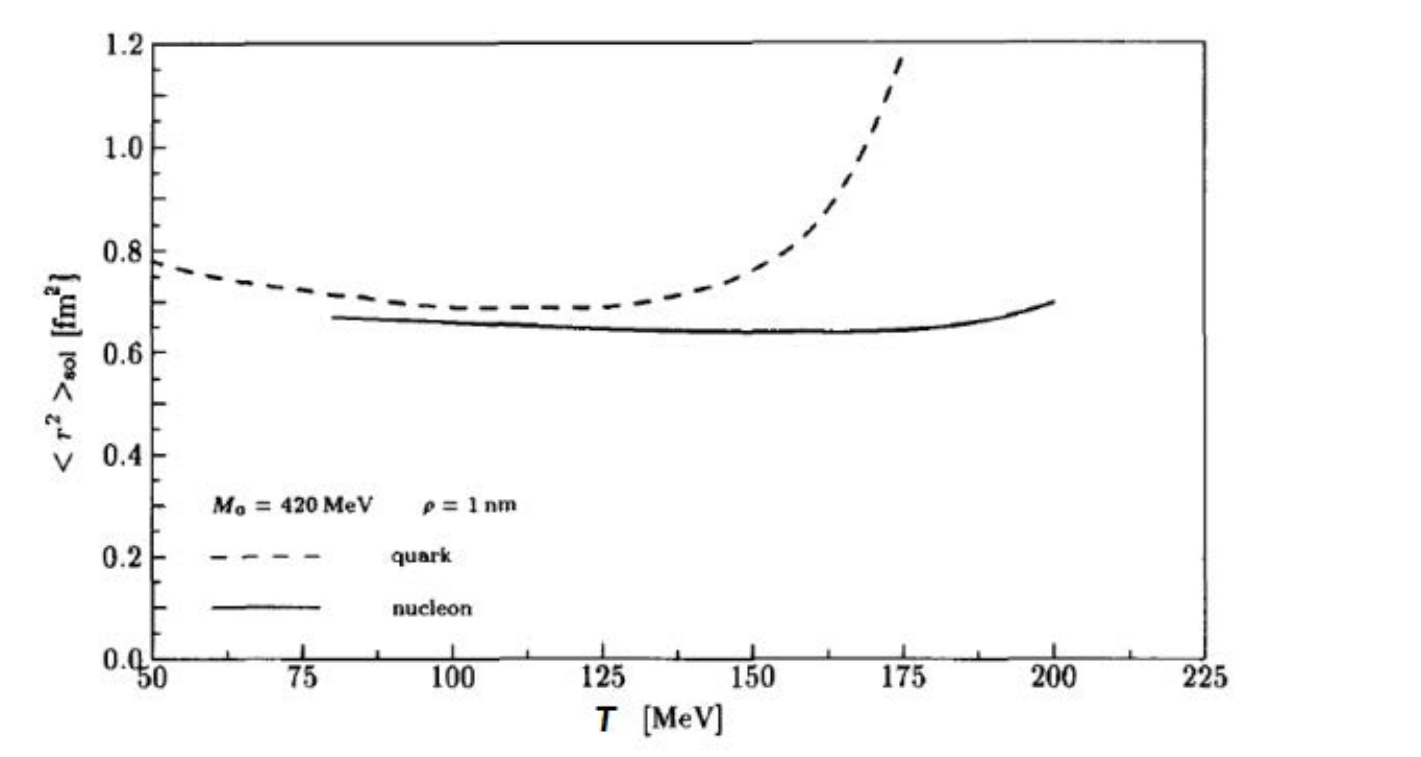}
\caption[]{B = 1 soliton m.s. radius in nucleon matter at p = pnm compared to that in quark matter}\label{f09}
\end{figure}

The calculated m.s. soliton radius for both quark and nucleon matter is presented in Figure~\ref{f09}. Close to the critical temperature it starts to
grow but it is much less pronounced than in the case of quark matter.

It should be noted also that the coupling constant $g_N$ defined as

\begin{equation}
\label{myeq15}
g_N =  E_{sol} / M \, ,
\end{equation}

stays almost constant (see \cite{myref12}) which means that the relation (\ref{myeq15}) is a good approximation also in a hot medium.

\section{Conclusion}

The bulk thermodynamic characteristics, meson properties and the properties of the nucleon as a $B = 1$ soliton of $N_c$ valence quarks with Dirac and Fermi seas are studied in a hot medium in the framework of an effective quark-meson theory. The Fermi sea of nucleons as well as of quarks is considered. At some critical values of temperature and/or density chiral phase transitions from the Goldstone to the Wigner phase are suggested. The quark matter is quite soft against thermal fluctuations whereas the nucleon matter is much stiffer and the corresponding phase transition is rather sharp. In the quark matter at low temperatures (below $90$ MeV) a first-order phase transition is expected. At higher temperatures it changes to a second-order phase transition. In contrast to the quark matter in the case of nucleon matter the model suggests a first-order phase transition even in case of vanishing density and high temperature. According to this model picture at higher temperatures a mix of different phases with different order parameters is expected. The bulk thermodynamic characteristics in nucleon and quark matter are different and at least according to the present results one has to consider also the baryon degrees of freedom in order to get a complete picture.   
 
The nucleon as a $B = 1$ soliton in medium is getting swelled and its mass is reduced as well.  In the baryon medium the soliton is less affected by the medium. At finite density the temperature stabilizes the soliton. At some critical values of temperature and/or density the nucleon as a soliton disappears. In the model this delocalization means a transition from nucleon to quark matter. The critical values for the delocalization of the soliton are  same as for the phase transition from nucleons to quarks in the nucleon matter which makes this model picture rather consistent.

\section*{Acknowledgements}
The author is greatful for the financial support of the Bulgarian Science Fund under 
Contract No. DFNI–T02/19.

\end{document}